\documentclass[11pt,a4paper]{article}

\usepackage{graphicx}
\usepackage{color}
\usepackage{amsmath,amssymb}
\usepackage{bbold}
\usepackage{t1enc}
\usepackage{cite}

\DeclareMathSymbol{\shortminus}{\mathbin}{AMSa}{"39}
\DeclareMathSymbol{\shm}{\mathbin}{AMSa}{"39}

\newcommand{\UU}{|{ \small ++} \rangle}
\newcommand{\UD}{|{ \small +-} \rangle}
\newcommand{\DU}{|{ \small -+} \rangle}
\newcommand{\DD}{|{ \small --} \rangle}
\newcommand{\U}{|{ \small +} \rangle}
\newcommand{\D}{|{ \small -} \rangle}

\begin{document}

\begin{center}
\begin{Large}
{\bf Decay of entangled fermion pairs with post-selection}
\end{Large}

\vspace{0.5cm}
\renewcommand*{\thefootnote}{\fnsymbol{footnote}}
\setcounter{footnote}{0}
J.~A.~Aguilar--Saavedra \\[1mm]
\begin{small}
Instituto de F\'isica Te\'orica IFT-UAM/CSIC, c/Nicol\'as Cabrera 13--15, 28049 Madrid, Spain \\
\end{small}
\end{center}

\begin{abstract}
We consider a pair of unstable fermions in a spin-entangled state.  After the decay of one fermion, a spin measurement is performed on the surviving partner, with a Stern-Gerlach experiment or similar.
The measurement not only projects the spin of the surviving fermion, but is also physically equivalent to a spin projection for the decayed one  --- even when it no longer exists. This post-selection effect would be experimentally accessible using muon pairs in a maximally-entangled state, produced either in the decay of a scalar particle, or in $e^+ e^-$ collisions at wide angles.
\end{abstract}

\section{Introduction}

Entanglement is one of the most striking features of quantum mechanics, 
as stressed by Schr\"odinger~\cite{sch1935}, 
and elementary particle physics offers many possibilities for novel tests.
Spin entanglement tests have been proposed for top quark-antiquark pairs, weak boson pairs, and between top quarks and $W$ bosons (see Ref.~\cite{Aguilar-Saavedra:2023hss} and references therein). The extremely short lifetime of these particles, of the order of $10^{-24}$ s, prevents any kind of direct spin measurement. However, because angular momentum is conserved in the decay, the spin state of the parent particles leaves its imprint in the angular distribution of their decay products. Therefore, given an ensamble of, say, $t \bar t$ pairs in a given spin state, the density operator of the pair can be reconstructed  from the expected value of observables involving the joint angular distribution of their decay products.\footnote{The experimental complications associated to the measurement, and the possible presence of backgrounds, are important issues but not relevant for the present discussion.}

Particle decay (as well as time evolution) is a component that is not present in entanglement experiments with electrons and photons, and leads to unique phenomena.
Let us consider fermion pairs $f_A f_B$ that are produced in a spin-entangled state described by the density operator $\rho_{AB}$, these fermions being sufficiently long-lived to be able to perform Stern-Gerlach (SG) experiments before they decay. If we perform a SG measurement on particle $A$ that gives a result $s= \pm 1/2$, the spin density operator of $B$ will be projected depending on the outcome of the experiment on $A$,
\begin{equation}
\rho_B = \frac{1}{p_s} \mathcal{P}_s^A \rho_{AB} \mathcal{P}_s^A  \,, \quad p_s = \operatorname{tr} [ \rho_{AB} \mathcal{P}_s^A ] \,,
\label{ec:rhoA}
\end{equation}
with $\mathcal{P}_s^A$ the projector onto the subspace of spin $s$ for particle $A$, and $p_s$ the probability to obtain this result. When particle $B$ subsequently decays, its decay products will {\em statistically} follow an angular distribution that precisely corresponds to the operator $\rho_B$ above, selected by the SG measurement on $A$.\footnote{Note that event by event, it not possible to determine the polarisation of $B$ from the kinematical configuration of its decay products.} Namely, if we perform the experiment on an ensemble of identical $f_A f_B$ pairs, selecting those pairs for which a specific value $s$ is obtained in the SG measurement, the angular distribution of the decay products of $B$ will follow the pattern that corresponds to the density operator (\ref{ec:rhoA}).

But, what if the SG measurement on $A$ is performed {\em after} the decay of $B$? In terms of spin, the decay of a particle is not in general a `measurement' that makes the system jump into a definite state. An experimental demonstration of this fact may eventually be provided by the post-decay entanglement measurements proposed in Ref.~\cite{Aguilar-Saavedra:2023hss}.
Then, if the first decay does not collapse the entangled state, it is the later SG measurement on $A$ that projects the spin state of the surviving partner --- and for particle $B$, it is physically equivalent to projecting the spin state before its decay. It is worth clarifying here that the probability to obtain either result for $A$ depends itself on the kinematics of the $B$ decay (this point will be further discussed in the following). To illustrate this point, let us consider a {\em gedankenexperiment} with $A$ and $B$ in a spin-singlet state,
\begin{equation}
|\Psi \rangle = \frac{1}{\sqrt 2} \left[ \U_A \D_B  - \D_A \U_B  \right] \,,
\end{equation}
where we label the spin-up and spin-down states as $\U$ and $\D$, as usual. If $B$ decay products are in an idealised kinematical configuration that univocally corresponds to the $\U$ state (in which case the decay would constitute a spin measurement), the SG measurement on $A$ would certainly obtain the value $\D$. But, despite this partial influence of the prior $B$ decay on the later SG result, the post-selection is quite intriguing. 
A previous post-tag proposal has been made for flavour entanglement in the $K^0 - \bar K^0$ system~\cite{Bernabeu:2019gjs}. In that case, however, there are other ingredients that complicate the interpretation, such as time evolution and kaon mixing, and the fact that kaon decays filter the state of the entangled pair.

In order to perform experimental tests of this post-selection one needs to consider a fermion with a lifetime large enough to allow it to travel macroscopic distances before decaying. The muon, with a mass $m_\mu = 105$ MeV and a lifetime of $2.2 \times 10^{-6}$ s at rest, seems a suitable candidate, even without significant time dilation effects to increase its decay length. For example, with an energy of twice its mass, the mean decay length is around 1.1 Km. (The $\tau$ lepton has a much shorter lifetime of $2.9 \times 10^{-13}$ s, and even if produced at multi-TeV energies, the decay length is of the order of 10 cm.) Muons are highly penetrating through matter, which is also an advantage to minimise possible decoherence effects. 

In the following we describe how this post-selection could be experimentally probed. In sections~\ref{sec:2} and~\ref{sec:3} we calculate in turn the density operator for $\mu^+ \mu^-$ pairs resulting from the decay of a scalar, and for $\mu^+ \mu^-$ pairs produced in low-energy electron-positron collisions, identifying the kinematical setups that provide maximally-entangled states and would allow the post-selection to be observed. We describe in section~\ref{sec:4} how the post-selection could be observed in muon decay distributions. The violation of Bell inequalities~\cite{Bell:1964kc} for the entangled dimuon pairs are addressed in section~\ref{sec:B}, and our results are discussed in section~\ref{sec:5}.

\section{Density operator for $S \to  \mu^+ \mu^-$}
\label{sec:2}

We consider a scalar $S$ of mass $M$ and generic interaction with the muon
\begin{equation}
\mathcal{L} = g \bar \mu (\cos \delta + i \sin \delta \gamma_5) \mu  S \,.
\end{equation}
The constant $g$ measures the overall strength of the interaction, and $\delta \in ]-\pi/2,\pi/2]$ is a phase that determines its parity properties, e.g. for $\delta = 0$ the interaction is of scalar type and for $\delta = \pi/2$ it is pseudo-scalar. 

We take the $\hat z$ axis as the spin quantisation direction.
In the $S$ rest frame, which coincides with the dimuon centre-of-mass (c.m.) frame, the momentum of the negative muon can be taken without loss of generality in the $xz$ plane,
\begin{equation}
p_1 = (M/2, k \sin \theta, 0, k \cos \theta) \,,
\end{equation}
with $k=[M^2/4 - m_\mu^2]^{1/2}$;
the positive muon has the same energy and three-momentum $\vec p_2 = - \vec p_1$. 
The polarised amplitudes for $S \to \mu^- (p_1,s_1) \mu^+ (p_2,s_2)$ can easily be calculated by using the explicit representation of the spinors, and from these the density operator is obtained. 
For the Hilbert space $\mathcal{H}_A \otimes \mathcal{H}_B$ of the muon pair spins we use the basis
\begin{equation}
\{ \UU \,, \UD \,, \DU \,, \DD  \} \,,
\label{ec:basis}
\end{equation}
where the first particle is $\mu^-$ and the second particle $\mu^+$. The spin density operator in this basis has matrix elements
\begin{align}
& \rho_{11} = \rho_{44} = -\rho_{14} = \mathcal{N}^{-1} \left[ 4 k^2 \cos^2 \delta \sin^2 \theta  \right]  \,, \notag \\
& \rho_{22} = \rho_{33} = \mathcal{N}^{-1} \left[ 4 k^2 \cos^2 \delta \cos^2 \theta + M^2 \sin^2 \delta  \right]  \,, \notag \\
& \rho_{12} = \rho_{13}^* = - \rho_{24}^* = -\rho_{34} = \mathcal{N}^{-1} \left[ 2 k^2 \cos^2 \delta \sin 2\theta + i M k \sin 2\delta \sin \theta \right] \,, \notag \\
%
%
%
& \rho_{23} =  \mathcal{N}^{-1} \left[ 2 k \cos \delta \cos \theta - i M \sin \delta \right]^2 \,,
%
%
%
\label{ec:rhoS}
\end{align}
with $\rho_{ij}^* = \rho_{ji}$ and
\begin{equation}
\mathcal{N} = 8 k^2 \cos^2 \delta + 2 M^2 \sin^2 \delta \,.
\end{equation}
Notice that we are not neglecting the muon mass, therefore the density operator given by (\ref{ec:rhoS}) describes the decay of any scalar into a pair of fermions with equal mass.
Setting $\theta = 0$ we recover the density operator calculated for Higgs boson decays $H \to \tau^+ \tau^-$ in Ref.~\cite{Altakach:2022ywa}. In our case, since a SG experiment is better suited to measure spin in a direction perpendicular to the particle motion, we set $\theta = \pi/2$. The density operator describes a pure state for all values of $\delta$. In the simplest cases of $J^P = 0^+$ ($\delta = 0$) and  $J^P = 0^-$ ($\delta = \pi/2$) we have
\begin{equation}
\rho_{0^+} = \frac{1}{2}
\left( \! \begin{array}{cccc}
1 & 0 & 0 & -1 \\
0 & 0 & 0 & 0 \\
0 & 0 & 0 & 0 \\
-1 & 0 & 0 & 1 \\
\end{array} \! \right)  \,, \quad
\rho_{0^-} = \frac{1}{2}
\left( \! \begin{array}{cccc}
0 & 0 & 0 & 0 \\
0 & 1 & -1 & 0 \\
0 & -1 & 1 & 0 \\
0 & 0 & 0 & 0 \\
\end{array} \! \right) \,.
\end{equation}
These operators correspond to maximally-entangled states
\begin{equation}
|\psi_{0^+} \rangle = \frac{1}{\sqrt 2} \left[ \UU - \DD \right] \,, \quad
|\psi_{0^-} \rangle = \frac{1}{\sqrt 2} \left[ \UD - \DU \right] \,.
\end{equation}
These results can be understood as follows~\cite{Altakach:2022ywa}. The parity-conserving decay of a $0^+$ scalar into a $\mu^+ \mu^-$  pair necessarily leads to an state of orbital angular momentum $L=1$ and total spin $S=1$, and the third component of the orbital angular momentum vanishes in the direction of motion of the particles, which we have taken in the $\hat x$ axis. On the other hand, the parity-conserving decay of a $0^-$ scalar produces $L = 0$, $S=0$, and the muon pair is in a spin-singlet state.

\section{Density operator for $e^+ e^- \to  \mu^+ \mu^-$}
\label{sec:3}

Muon pairs can be produced in the Drell-Yan process mediated by a virtual photon and $Z$ boson.
At GeV-scale energies well below the $Z$-boson threshold, its contribution can be ignored in order to simplify the computations. We also neglect the electron mass.
The $\hat x$ axis is taken in the beam direction, and the $\hat y$ axis perpendicular to it, in the production plane. In the c.m. frame the momenta of the incident electron and the outgoing negative muon are, respectively,
\begin{equation}
p_3 = (E/2,E/2,0,0) \,,\quad p_1 = (E/2,k \cos \theta,k\sin \theta,0) \,,
\end{equation}
with $E$ being the c.m. energy of the collision, and $k=[E^2/4 - m_\mu^2]^{1/2}$.
The positron and positive muon have momenta $p_4 = (E/2,-\vec p_3)$, $p_2 = (E/2,-\vec p_1)$, respectively. 
We take the $\hat z$ axis as spin quantisation direction. The amplitudes averaged over $e^+$, $e^-$ spins can easily be calculated by using the explicit representation of the spinors in the Dirac basis.
From the polarised amplitudes the density operator for the muon pair is obtained. In the basis (\ref{ec:basis}), it has matrix elements
\begin{align}
& \rho_{11} = \rho_{44} = \mathcal{N}^{-1} \left[ E^2 + 4 m_\mu^2 + (E^2 - 4 m_\mu^2) \cos 2 \theta \right] \,, \notag \\
& \rho_{22} = \rho_{33} = \rho_{23} = \mathcal{N}^{-1} \left[ 2 E^2 \right] \,, \notag \\
& \rho_{12} = \rho_{13} = \rho_{24} = \rho_{34} = \mathcal{N}^{-1} \left[ 2 E \, e^{-i  \theta} (E \cos \theta + i \, 2 m_\mu \sin \theta)  \right] \,, \notag \\
& \rho_{14} =  \mathcal{N}^{-1} \left[ 2 e^{-i 2 \theta} (E \cos \theta + i \, 2 m_\mu \sin \theta)^2 \right] \,,
\end{align}
with $\rho_{ij}^* = \rho_{ji}$ and
\begin{equation}
\mathcal{N} = 6 E^2 + 8 m_\mu^2 + 2 (E^2 -4 m_\mu^2) \cos 2 \theta \,.
\end{equation}
For $\theta = \pi/2$ the density operator is
\begin{equation}
\rho = \frac{1}{2 E^2 + 8 m_\mu^2}
\left( \! \begin{array}{cccc}
4 m_\mu^2 & 2 E m_\mu & 2 E m_\mu & 4 m_\mu^2 \\
2 E m_\mu & E^2 & E^2 & 2 E m_\mu \\
2 E m_\mu & E^2 & E^2 & 2 E m_\mu \\
4 m_\mu^2 & 2 E m_\mu & 2 E m_\mu & 4 m_\mu^2 \\
\end{array} \! \right) \,.
\end{equation}
It corresponds to the pure state
\begin{equation}
|\psi \rangle = \frac{1}{[2 E^2 + 8 m_\mu^2]^{1/2}} \left[ E ( \UD + \DU ) + 2 m_\mu ( \UU + \DD ) \right] \,,
\end{equation}
which for $E \gg m_\mu$ is the $|10\rangle$ triplet state with vanishing third component. A similar result was previously obtained for ultrarrelativistic top quarks produced in $q \bar q$ collisions~\cite{Mahlon:1997uc}.

\section{Muon post-selection}
\label{sec:4}

Muons decay in the channel $\mu^\pm \to e^\pm \nu_e  \nu_\mu$ mediated by an off-shell $W$ boson.
The two neutrinos leave the detectors unobserved, and one can only access kinematical variables for the emitted $e^\pm$, which we take massless since $m_e / m_\mu \sim 1/200$. The normalised decay rate in the muon rest frame is, after integrating over the momenta of the two  neutrinos~\cite{Bouchiat:1957zz}
\begin{equation}
\frac{1}{\Gamma} \frac{d\Gamma}{dx \, d\!\cos \theta_{e}^*} = x^2 \left[ (3 - 2 x) + (2 x - 1) \kappa P_\mu \cos \theta_{e}^* \right] \,,
\label{ec:mudec}
\end{equation}
where $x = E_e / E_e^\text{max}$ is the $e^\pm$ energy normalised to its maximum value $E_e^\text{max} = m_\mu/2$; $\theta_e^*$ is the angle between the $e^\pm$ three-momentum and an arbitrary direction $\hat n$; $P_\mu$ is the muon polarisation in that axis, $P_\mu = 2 \langle \vec S \cdot \hat n \rangle$; $\kappa = 1$ for $\mu^+$ decays and $\kappa = -1$ for $\mu^-$ decays. Integrating over $e^\pm$ energies,
\begin{equation}
\frac{1}{\Gamma}\frac{d\Gamma}{d\!\cos \theta_e^*} = \frac{1}{2} \left[ 1 + \frac{1}{3} \kappa P_\mu \cos \theta_e^* \right] \,.
\label{ec:mudec1D}
\end{equation}
Let us focus on the two examples presented in sections~\ref{sec:2} and \ref{sec:3}, and the decay of the positive muon ($\kappa = 1$). The spin quantisation axis is the $\hat z$ direction ($\hat n = \hat z$) and the muon momenta in the c.m. frame lie in the $xy$ plane. The positrons are then produced in the muon rest frame with an up-down asymmetry 
\begin{equation}
\int_0^1 \frac{1}{\Gamma}\frac{d\Gamma}{d\!\cos \theta_{e^+}^*}
- \int_{-1}^0 \frac{1}{\Gamma}\frac{d\Gamma}{d\!\cos \theta_{e^+}^*} = \frac{P_\mu}{6} \,.
\label{ec:AP}
\end{equation}
Moreover, because the $\mu^+$ momentum is perpendicular to the $\hat z$ axis, the up-down asymmetry is maintained in the c.m. frame. For $\U$ and $\D$ states ($P_\mu = \pm 1$), the up-down asymmetry is $\pm 1/6$, respectively. The actual shape of the $\cos \theta_{e^+}$ distribution --- where the polar angle $\theta_{e^+}$ with respect to the $\hat z$ axis is measured in the c.m. frame --- depends on the muon energy. It can be numerically obtained by generating a random sample of $e^+$ momenta in the $\mu^+$ rest frame that obey the distribution (\ref{ec:mudec}) and subsequently boosting the momenta to the c.m. frame.

We are now in position to discuss the post-selection on $\mu^+$ decay distributions. We consider for definiteness the dimuon pairs resulting from the decay of a $\eta$ meson, which is a $J^P = 0^-$ scalar with a mass $m_\eta = 548$ MeV. (A similar reckoning can be made for Drell-Yan production at $\theta = \pi/2$.) As shown in section~\ref{sec:2}, for this example the muons are in a spin-singlet state. For a given kinematical configuration $|f\rangle$ in the $\mu^+$ decay, the state of the negative muon is $ \left[a_- \U - a_+ \D \right] $,
with $a_\pm = \langle f | T | {\small \pm} \rangle$ the decay amplitudes for the positive muon in $\U$, $\D$ states, suitably normalised so that $|a_+|^2+|a_-|^2 = 1$.\footnote{The same happens with the post-decay entanglement discussed in Ref.~\cite{Aguilar-Saavedra:2023hss}, see also Ref.~\cite{Bernabeu:2019gjs}.}
A SG measurement on $\mu^-$ then obtains $\U$ with probability $|a_-|^2$, and $\D$ with probability $|a_+|^2$. 

\begin{figure}[htb]
\begin{center}
\includegraphics[height=5.5cm,clip=]{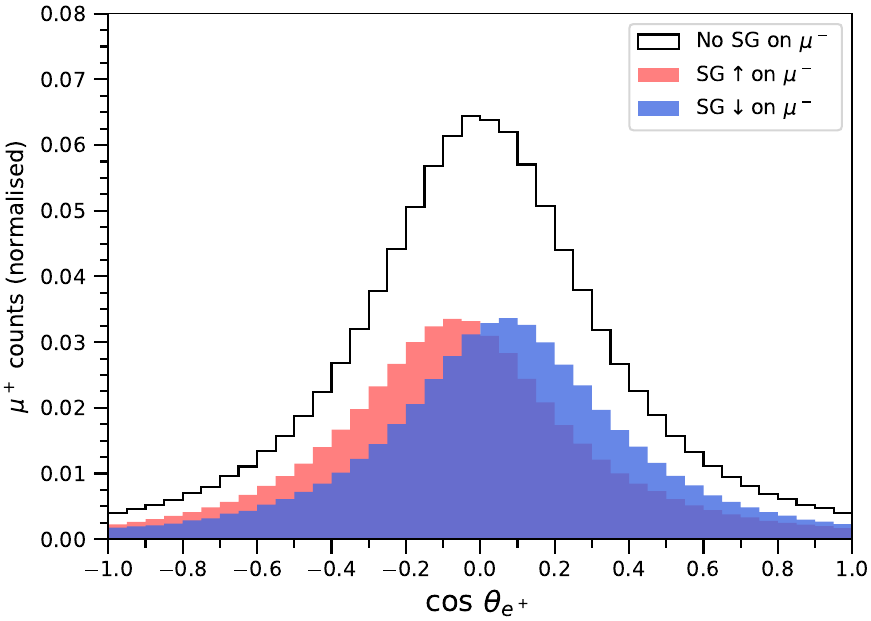} 
\caption{Polar angle distribution for the $e^+$ resulting from $\mu^+$ decay with post-selection (red and blue lines) and without any measurement on $\mu^-$ (black line). See the text for details.}
\label{fig:costh}
\end{center}
\end{figure}

For simplicity we concentrate on the $\cos \theta_{e^+}$ distribution in the $\eta$ rest frame, although multi-dimensional distributions involving the energy and/or azimuthal angle of the positron could be studied as well. 
The $\cos \theta_{e^+}$ distributions for either outcome of the SG measurement are presented in Fig.~\ref{fig:costh}. In case no measurement is performed on $\mu^-$, the symmetric black line is obtained. Notice also that a random selection of events would not produce any asymmetry in $\cos \theta_{e^+}$: the asymmetry is precisely due to the $\mu^+ \mu^-$ entanglement.

As shown in section~\ref{sec:3}, results for $e^+ e^-$ scattering at $\theta = \pi/2$ are alike for GeV-scale energies for which the muon mass is negligible. The c.m. up-down asymmetry is $\pm 1/6$ in all cases but with increasing energy the distributions are more peaked near $\cos \theta_{e^+} = 0$ due to the Lorentz boost from the $\mu^+$ rest frame to the c.m. frame. For the decay of a parity-even scalar the blue and red areas are swapped.

\section{Bell inequalities}
\label{sec:B}

It is easy to verify that for the maximally-entangled dimuon pairs generated in the processes discussed above, the Clause-Horne- Shimony-Holt (CHSH) inequalities~\cite{Clauser:1969ny} are violated. These inequalities read
\begin{equation}
\left| \langle AB \rangle - \langle AB' \rangle + \langle A'B \rangle + \langle A'B' \rangle \right| \leq 2 \,,
\label{ec:CHSH}
\end{equation}
where $A$, $A'$ are two operators for $\mu^-$ (Alice) and $B$, $B'$ spin operators for $\mu^+$ (Bob), normalised to have eigenvalues $\pm 1$. We again take as example the spin-singlet state. We can choose
\begin{align}
& A = \sigma_3 \,,\quad A' = \sigma_2 \,,\quad B = \frac{1}{\sqrt 2}(\sigma_2 + \sigma_3) \,,\quad
B' = \frac{1}{\sqrt 2}(\sigma_2 - \sigma_3) \,,
\end{align}
with $\sigma_i = 2 S_i$ the Pauli matrices. For the spin-singlet state,
\begin{equation}
\langle AB \rangle =  - \langle AB' \rangle = \langle A'B \rangle = \langle A'B' \rangle = -\frac{1}{\sqrt 2} \,,
\end{equation}
so the l.h.s. of (\ref{ec:CHSH}) equals $2\sqrt 2$, violating the inequality. Experimentally, this could be done as follows.

Bob can only measure polarisations by means of asymmetries, c.f. (\ref{ec:AP}). In the $yz$ plane, $\langle B \rangle$ is measured from the asymmetry along the bisector of the second and fourth quadrants, with the positive direction in the first quadrant. Likewise, $\langle B' \rangle$ is measured from the asymmetry along the bisector of the first and third quadrants, with the positive direction in the fourth quadrant. The expected values are six times the measured value of the asymmetry, as it follows from (\ref{ec:AP}).

For each dimuon pair, and after Bob registers the $\mu^+$ decay, Alice chooses whether to measure spin in the $\hat z$ or $\hat y$ direction for $\mu^-$. The expected values are computed by weighing 
$\langle B \rangle$ and $\langle B' \rangle$ with the eigenvalue of $A$ or $A'$, for example
\begin{equation}
\langle AB \rangle = \frac{1}{2} \left[ \langle B \rangle_\uparrow - \langle B \rangle_\downarrow \right] \,,
\end{equation}
where the up (down) arrows indicate that the averages are restricted to events with positive (negative) eigenvalue obtained in the measurement of $S_3$ for the negative muon. We stress that post-selection is at work here too: when $S_3$ is measured for the negative muon, $\langle B \rangle_\uparrow = - \langle B \rangle_\downarrow = -1/\sqrt{2}$ for $\mu^+$. Similar remarks apply to the remaining expected values.

\section{Discussion}
\label{sec:5}

The first point raised by these results is whether causality would be violated by post-selection. Obviously it is not. Once $\mu^+$ decays into $e^+ \nu_e \bar \nu_\mu$ of given momenta (and spins, which are not measured), these are not changed by what we do with $\mu^-$. Post-selection is rather a splitting of a sample of already `recorded' $\mu^+$ events, based on what we later measure on $\mu^-$ --- noting that the outcomes of the SG measurement have different probabilities depending on the $\mu^+$ decay kinematics, as discussed above.

That being said, the effect is non-trivial and its observation would be an important new test of quantum mechanics. The SG measurement on $\mu^-$ indeed collapses its spin state and is physically equivalent to projecting the $\mu^+$ spin into a definite state. In the example $\eta \to \mu^+ \mu^-$ of the preceding section, with muons moving in the $\hat x$ axis, we have chosen to measure spin in the $\hat z$ axis. We could also choose to make a SG measurement in the $\hat y$ axis in which the spins are fully anticorrelated too, and we would obtain a front-rear asymmetry, instead of up-down. However, if we measured spin in the $\hat y$ axis after measuring it in the $\hat z$ axis, we would not observe any asymmetry: the spin state would already have collapsed. We can even make alternate measurements, as described in section~\ref{sec:B}, to test the violation of the CHSH inequalities for the entangled dimuon pair.

Finally, from a practical point of view we note that there are not many scalar particles with $J^P = 0^-$ that can allow to perform this type of experiments, but Drell-Yan production at $\theta = \pi/2$ is perfectly suited as well. Relatively low energies are preferred in order to avoid large time-dilation effects that would require a large experimental apparatus to observe the $\mu^+$ decay. Stern-Gerlach experiments are more common with neutral particles but they can also be performed with charged ones, as long as an electric field is also introduced to compensate for the Lorentz force due to the magnetic field.
Also, the timing of the $\mu^+$ decay and the SG measurement on $\mu^-$ would have to be carefully measured, in order to ensure that it is a post-selection effect and not a pre-selection. In principle, this could be done by adjusting the distance of the $\mu^+$ detector to the source.
The specific details of possible experimental implementations are out of the scope of this work.

\section*{Acknowledgements}

I thank J. Bernab\'eu for useful discussions about Ref.~\cite{Bernabeu:2019gjs}, and the CERN TH Department for hospitality during the realisation of this work. This work of has been supported by MICINN projects PID2019-110058GB-C21, PID2022-142545NB-C21 and CEX2020-001007-S funded by MCIN/AEI/10. 13039/501100011033 and by ERDF, and by Funda\c{c}{\~a}o para a Ci{\^e}ncia e a Tecnologia (FCT, Portugal) through the project CERN/FIS-PAR/0019/2021.


\begin{thebibliography}{99}

\bibitem{sch1935}
E. Schrödinger, ``Discussion of Probability Relations between Separated Systems'',
Math. Proc. Cambridge Philos. Soc. 31, 555 (1935).



\bibitem{Aguilar-Saavedra:2023hss}
J.~A.~Aguilar-Saavedra,
``Post-decay quantum entanglement in top pair production,''
Phys. Rev. D \textbf{108}, no.7, 076025 (2023)
[arXiv:2307.06991 [hep-ph]].


\bibitem{Bernabeu:2019gjs}
J.~Bernab\'eu and A.~Di Domenico,
``Can future observation of the living partner post-tag the past decayed state in entangled neutral K mesons?,''
Phys. Rev. D \textbf{105}, no.11, 116004 (2022)
[arXiv:1912.04798 [quant-ph]].

\bibitem{Bell:1964kc}
J.~S.~Bell,
``On the Einstein-Podolsky-Rosen paradox,''
Physics Physique Fizika \textbf{1}, 195-200 (1964)

\bibitem{Altakach:2022ywa}
M.~M.~Altakach, P.~Lamba, F.~Maltoni, K.~Mawatari and K.~Sakurai,
``Quantum information and CP measurement in $H \to \tau^+ \tau^-$ at future lepton colliders,''
Phys. Rev. D \textbf{107}, no.9, 093002 (2023)
[arXiv:2211.10513 [hep-ph]].

\bibitem{Mahlon:1997uc}
G.~Mahlon and S.~J.~Parke,
``Maximizing spin correlations in top quark pair production at the Tevatron,''
Phys. Lett. B \textbf{411}, 173-179 (1997)
[arXiv:hep-ph/9706304 [hep-ph]].

\bibitem{Bouchiat:1957zz}
C.~Bouchiat and L.~Michel,
``Theory of $\mu$-Meson Decay with the Hypothesis of Nonconservation of Parity,''
Phys. Rev. \textbf{106}, 170-172 (1957)

\bibitem{Clauser:1969ny}
J.~F.~Clauser, M.~A.~Horne, A.~Shimony and R.~A.~Holt,
``Proposed experiment to test local hidden variable theories,''
Phys. Rev. Lett. \textbf{23}, 880-884 (1969)

\end{thebibliography}
\end{document}